\documentclass[aps,pra]{revtex4}

\usepackage{graphicx}
\usepackage{amsmath}
\usepackage{amssymb}
\usepackage{latexsym}

\newcommand{\be}{\begin{equation}}
\newcommand{\ee}{\end{equation}}
\newcommand{\ba}{\begin{array}}
\newcommand{\ea}{\end{array}}
\newcommand{\calH}{{\cal H }}

\newcommand{\calM}{{\cal M }}

\newcommand{\calF}{{\cal F }}
\newcommand{\calG}{{\cal G }}

\newcommand{\CC}{\mathbb{C}}

\newcommand{\NP}{$\mathrm{NP}$}
\newcommand{\CP}{$\mathrm{C_{=}P}$}

\newcommand{\QMA}{$\mathrm{QMA}$}
\newcommand{\BQP}{$\mathrm{BQP}$}
\newcommand{\QMAe}{$\mathrm{QMA_1}$}
\newcommand{\la}{\langle}
\newcommand{\ra}{\rangle}
\newcommand{\nn}{\nonumber}

\newcommand{\trace}{\mathop{\mathrm{Tr}}\nolimits}

\newcommand{\ep}{\epsilon}

\newcommand{\nrb}[1]{\mathop{\mathrm{nbgh}(#1)}}
\newcommand{\clock}{\mbox{\scriptsize clock}}
\newcommand{\legal}{\mbox{\scriptsize legal}}
\newcommand{\init}{\mbox{\scriptsize init}}
\newcommand{\comp}{\mbox{\scriptsize comp}}
\newcommand{\prop}{\mbox{\scriptsize prop}}
\newcommand{\out}{\mbox{\scriptsize out}}

\newtheorem{definition}{Definition}
\newtheorem{lemma}{Lemma}

\begin{document}
\title{Efficient algorithm for a quantum analogue of $2$-SAT}

\author{Sergey Bravyi}

\affiliation{\mbox{IBM Watson Research Center, Yorktown Heights, NY 10598}}

\date{\today}

\begin{abstract}
Complexity of a quantum analogue of the satisfiability problem
is studied.
Quantum $k$-SAT is a problem of verifying
whether there exists a state $|\Psi\ra$ of $n$ qubits
such that its $k$-qubit reduced density matrices have support on
prescribed subspaces.
We present a classical algorithm solving quantum \mbox{$2$-SAT}
in a polynomial time. It generalizes the well-known
algorithm for the classical $2$-SAT.
Besides, we show that for any $k\ge 4$ quantum $k$-SAT is
complete in the complexity class
\QMA{} with one-sided error.
\end{abstract}

\maketitle

\section{Introduction}

Quantum analogues of classical complexity classes have been studied extensively
during the last several years.
 From the practical perspective the most interesting
of them is \BQP{} --- a class of problems that can be solved on a quantum computer
in a polynomial time with bounded error probability.
Another relevant class is \QMA{} --- a quantum analogue of \NP. By definition, a language
$L$ belongs to \QMA{} if a membership $x\in L$ can be efficiently verified on a quantum
computer having access to a witness quantum state $|\psi_x\ra$ playing the same role as
a witness bit string in the definition of \NP, see~\cite{KSV}.
It is widely believed that \QMA{} is strictly larger than \NP.

There are only a few natural problems known to be \QMA-complete.
The results of the present paper concern the problem $2$-local Hamiltonian
introduced by Kitaev, see~\cite{KKR,KSV}. It is a suitably formalized version
of a fundamental problem in the quantum many-body physics: evaluate the ground state
energy of $n$-qubit Hamiltonian which can be represented a sum of
two-qubit interactions, $H=-\sum_{a<b} H_{a,b}$.

A theorem proven by Kitaev, Kempe, and Regev at~\cite{KKR} asserts that
the problem $2$-local Hamiltonian is \QMA-complete.
One way to interpret this result is to identify a classical analogue of
the $2$-local Hamiltonian problem. If one assumes that all interactions $H_{a,b}$
are diagonal in the standard $|0\ra,|1\ra$ basis of $n$ qubits and that
their eigenvalues may be either $0$ or $1$, the $2$-local Hamiltonian problem
becomes equivalent to MAX-$2$-SAT~\cite{KKR}.
Indeed, in this case a ground state of $H$ can be chosen as a basis vector $|x\ra$
corresponding to some $n$-bit string $x$.
An eigenvalue of $H_{a,b}$ on $|x\ra$
is a Boolean function of bits $x_a, x_b$ that can be represented
as a conjuction of clauses $x_a\vee x_b$, $(\neg x_a)\vee x_b$,
$x_a\vee (\neg x_b)$, $(\neg x_a) \vee (\neg x_b)$
(non-trivial interactions involve one, two, or three clauses).
Accordingly, one can replace $H$ by a list of $2$-bit clauses, while
the ground state energy of $H$ is equal to the maximum number of clauses that
can be satisfied simultaneously.

This quantum-to-classical mapping allows one to attribute \QMA-completeness of
the $2$-local Hamiltonian problem to \NP-completeness of MAX-$2$-SAT.
On the other hand, classical problem $2$-SAT (verify that {\it all}
$2$-bit clauses in a given list can be satisfied simultaneously)
can be solved in a linear time~\cite{APT}, see also
Appendix~B.
It rises the following natural questions:
\begin{itemize}
\item  What is a quantum analogue of  $2$-SAT ?
\item Does there exist a poly-time algorithm (classical or quantum) solving
quantum $2$-SAT ?
\end{itemize}

To define quantum $2$-SAT we shall substitute a
$2$-bit clause $C_{a,b}$ involving some particular pair of bits $x_a,x_b$
(for example, $C_{a,b}=x_a\vee (\neg x_b)$) by a pure two qubit
state $|\phi_{a,b}\ra\in \CC^2\otimes \CC^2$.
Quantum analogue of a property ``$n$-bit string $x$ satisfies the clause $C_{a,b}$''
is a property ``quantum state $|\Psi\ra$
of $n$ qubits has reduced density matrix $\rho_{a,b}$ orthogonal to $|\phi_{a,b}\ra$''.
The latter means precisely that $\la \phi_{a,b}|\rho_{a,b}|\phi_{a,b}\ra=0$,
where $\rho_{a,b}=\trace_{c\ne a,b} |\Psi\ra\la \Psi|$.
Such a generalization seems justified since a two-bit clause $C_{a,b}$
is {\it not} satisfied only for one configuration of variables
$x_a,x_b$ (in the example above $x_a=0$ and $x_b=1$), so a satisfying
assignment $x$ must be orthogonal to this configuration.
In general, there might be several clauses involving a pair of bits $a,b$.
The corresponding two-qubit states $|\phi_{a,b}\ra$ span a
{\it forbidden subspace} $\calM_{a,b}\subseteq \CC^2\otimes \CC^2$,
and a satisfying assignment $|\Psi\ra$ must satisfy
$\la \phi|\rho_{a,b}|\phi\ra=0$ for any $|\phi\ra\in \calM_{a,b}$.
Denoting $\Pi_{a,b}$ an orthogonal projector onto $\calM_{a,b}$
multiplied by the identity operator on all other qubits,
we obtain an equivalent condition $\Pi_{a,b}\, |\Psi\ra=0$.
Thus quantum $2$-SAT can be defined as follows:

\vspace{5mm}

\noindent
{\bf Input:} {\it An integer $n$ and a family of $2$-qubit projectors
$\{\Pi_{a,b}\}$, $1\le a<b\le n$.}\\
{\bf Problem:} {\it Decide whether there exists a state $|\Psi\ra$
of $n$ qubits such that $\Pi_{a,b}\, |\Psi\ra=0$ for all
$1\le a<b\le n$.}

\vspace{5mm}

Note that there is no loss of generality in
imposing constraints for every pair of qubits, since some of the projectors
$\Pi_{a,b}$ may be equal to zero.

The main conclusion drawn in the present paper is that quantum $2$-SAT is
not harder than its  classical counterpart.  An efficient algorithm for
quantum $2$-SAT is described in Section~\ref{sec:2sat}.
The key ingredient of the algorithm is Lemma~\ref{lemma:key}
that allows one to generate new $2$-qubit constraints from existing ones
according to simple local rules. It can be illustrated by the following
simple example. Suppose a state $|\Psi\ra$
satisfies constraints
$\Pi_{1,2}\, |\Psi\ra=\Pi_{2,3}\, |\Psi\ra=0$,
where $\Pi_{1,2}=|\Psi^-\ra\la \Psi^-|_{1,2}\otimes I_3$,
and $\Pi_{2,3}=I_1\otimes |\Psi^-\ra\la \Psi^-|_{2,3}$
are projectors onto the singlet state, $|\Psi^-\ra=2^{-1/2}\, (|0,1\ra-|1,0\ra)$.
Taking into account that $I-\Pi_{1,2}$ and $I-\Pi_{2,3}$
are projectors onto the symmetric subspace of qubits
$1,2$ and $2,3$ respectively, we conclude that $|\Psi\ra$ is invariant under
swaps of qubits $1\leftrightarrow 2$ and $2\leftrightarrow 3$.
Thus $|\Psi\ra$ is also invariant under a swap $1\leftrightarrow 3$.
Therefore we can generate a new constraint $\Pi_{1,3}\, |\Psi\ra=0$,
where $\Pi_{1,3}=|\Psi^-\ra\la \Psi^-|_{1,3}\otimes I_2$.
Lemma~\ref{lemma:key} generalizes this observation to
projectors onto arbitrary two-qubit states (not necessarily entangled).

The algorithm is defined inductively, such that it allows one to reduce
$n$-qubit $2$-SAT to $(n-1)$-qubit $2$-SAT in a polynomial time.
Each step of the induction proceeds as follows (for a detailed
analysis see Section~\ref{sec:2sat}).
For each pair $1\le a<b\le n$ we look at the rank $r_{a,b}$ of the
projector $\Pi_{a,b}$
(the dimension of the forbidden subspace $\calM_{a,b}$).
If $r_{a,b}=4$, the problem has no satisfying
assignments. If $r_{a,b}=3$, a state of qubits $a,b$ is completely specified
by a constraint $\Pi_{a,b}\, |\Psi\ra=0$ alone, so one can exclude them
from consideration reducing $n$ to $n-2$. If $r_{a,b}=2$,
qubits $a$ and $b$ can be merged into a single logical qubit supported on
the zero subspace of $\Pi_{a,b}$. It reduces $n$ to $n-1$.
If $r_{a,b}\le 1$ for all $a$ and $b$,
we invoke  Lemma~\ref{lemma:key} to generate
new constraints. The process of generating new constraints
terminates whenever $r_{a,b}\ge 2$ is encountered at
some pair of qubits (which reduces the total number of qubits),
or if we arrive to what we call {\it a complete set of constraints}.
We prove then that any instance with a complete set of constraints
has a satisfying assignment
(moreover, it can be chosen as a product of one-qubit states).
The running time of the whole algorithm is $O(n^4)$.

A satisfiability problem in which clauses involve $k$ bits is known
as $k$-SAT. Quantum analogue of $k$-SAT can be specified by a family
of  $k$-qubit projectors $\{\Pi_S\}$, where $S\subseteq
\{1,\ldots,n\}$ runs over all subsets of cardinality $k$. One needs
to verify whether there exists an $n$-qubit state $|\Psi\ra$ such
that $\Pi_S\, |\Psi\ra=0$ for all $S$. Since $k$-SAT is known to be
\NP-complete for $k\ge 3$, one can ask whether quantum $k$-SAT is
\QMA-complete for $k\ge 3$? Before addressing this question it
should be noted that the definition of quantum $k$-SAT given above
is not quite satisfactory. Indeed, there might exist instances which
have no exact satisfying assignments, meanwhile having an approximate
solutions $|\Psi\ra$ such that equations $\Pi_S\, |\Psi\ra=0$ are
satisfied with an exponentially small error. To exclude such cases
from consideration we shall introduce a precision parameter $\ep$,
 $\ep\ge 1/n^\alpha$, $\alpha=O(1)$, separating positive and negative
instances.
The definition of quantum $k$-SAT given above must be complimented as
follows:

\vspace{5mm}

\noindent
{\bf Input:} {\it An integer $n$, a real number
$\ep=\Omega(1/n^\alpha)$,  and a family of $k$-qubit projectors
$\{\Pi_S\}$, $S\subseteq \{1,\ldots,n\}$, $|S|=k$. \\
{\bf Promise:} Either there exists $n$-qubit state $|\Psi\ra$
such that $\Pi_S\, |\Psi\ra=0$ for all $S$, or
$\sum_S \la \Psi|\Pi_S|\Psi\ra\ge \ep$ for all $|\Psi\ra$.\\
{\bf Problem:} Decide which one is the case.\\
}
\vspace{5mm}

To analyze complexity of quantum $k$-SAT we introduce a
class \QMAe.
It is defined in the same way as
\QMA{} with the only difference that for positive
instances  the verifying circuit  accepts with probability one,
see Section~\ref{sec:4sat} for more
details. The definition implies that \QMAe~$\subseteq$~\QMA.

We prove that quantum $k$-SAT with the promise as above is
\QMAe-complete for $k\ge 4$.
This result is quite unexpected since in general a subspace
spanned by satisfying assignments of quantum $k$-SAT lacks a
description in terms of $l$-qubit projectors for any $l<k$.  So one
could expect the complexity of quantum $k$-SAT to grow with $k$. Our
result shows that this is not the case, at least for $k\ge 4$.
Whether or not quantum $3$-SAT is \QMAe-complete remains an open
question.

The fact that quantum $5$-SAT is \QMAe-complete
easily follows from a possibility to represent the ``computational
history state'' associated with a quantum circuit as a common
zero vector of $5$-qubit projectors, see~\cite{KSV}.
A contribution reported in the present paper is essentially a reduction from $k=5$
to $k=4$.
A mapping from a quantum circuit to a family of $4$-qubit constraints
that we use
is a significantly simplified version
of a construction proposed in~\cite{ADKLLR}
(Section~4) for adiabatic quantum computation.

{\it Remark:}
Input data for all problems discussed in the paper involve linear operators
with complex matrix elements.
To deal with exact equalities one has to use an appropriate
exact representation of complex numbers.  A good choice is algebraic
numbers of bounded degree over the field of rational numbers
(roots of polynomials with rational coefficients).
All common linear algebra tasks for operators whose
matrix elements are algebraic numbers can be solved efficiently,
see books~\cite{AHU,BP94} for the subject.

The rest of the paper is organized as follows.
Efficient algorithm for quantum $2$-SAT is presented in
Section~\ref{sec:2sat} (for the sake of completeness we
outline the standard algorithm solving classical $2$-SAT in
Appendix~B). \QMAe-completeness of quantum $k$-SAT, $k\ge 4$,
is proved in Section~\ref{sec:4sat}.
A technical lemma needed for this proof concerning universality of
three-qubit quantum gates
with matrix elements from a fixed field is placed in
Appendix~A.

\section{Efficient algorithm for quantum $2$-SAT}
\label{sec:2sat}

Let $\{\Pi_{a,b}\}$, $1\le a<b\le n$ be an instance of quantum $2$-SAT
defined on $n$ qubits.
Without loss of generality $n\ge 3$.
Obviously, these data can be encoded by a binary
string $x$ of length $O(n^2)$.
We shall construct an algorithm that takes $x$ as input
and outputs one of the following
\begin{itemize}
\item{Output~1:} $x$ has no satisfying assignments.
\item{Output~2:} A list of one-qubit states $\{|\psi_j\ra\}$ such that
$|\psi_1\ra\otimes \cdots \otimes |\psi_n\ra$
is a satisfying assignment for $x$.
\item{Output~3:} An instance $y$ of quantum $2$-SAT defined on $n-1$
or smaller number of qubits, such that $y$ is
equivalent to $x$.
\end{itemize}
The algorithm runs in a time $O(n^3)$.
By applying it inductively $O(n)$ times one can reduce  $x$
to an equivalent instance involving a constant (say, $n=2$)
number of qubits which can be solved directly.
It should be noted that although the algorithm allows one to construct
one particular satisfying assignment (SA) for positive instances,
this SA might not be a product of one-qubit states
(the reduction corresponding to Output~3 may transform a product SA
for $y$ into entangled SA for $x$).
 Denote $r_{a,b}$ a rank of the projector $\Pi_{a,b}$.
Without loss of generality $0\le r_{a,b}\le 3$.
\begin{definition}
An instance of quantum $2$-SAT is called homogeneous iff
$r_{a,b}\le 1$ for all $a,b$.
\end{definition}
(Note that if $r_{a,b}=0$, there is no any constraint for a pair of qubits $a,b$.)
The first step of the algorithm is to verify whether $x$ is homogeneous.
If it is not, $x$ can be transformed into an equivalent instance $y$
on a smaller number of qubits (producing Output~3) as explained
in  Subsections~A,B. If $x$ is homogeneous (which is the most
interesting case) one has to proceed to Subsection~C.

\subsection{Projectors of rank three}
Suppose that $r_{a,b}=3$  for some pair of qubits $a,b$,
i.e.,
\[
\Pi_{a,b}=I-|\phi\ra\la\phi|, \quad |\phi\ra\in
\CC^2 \otimes \CC^2, \quad \la \phi|\phi\ra=1.
\]
An eigenvalue equation $\Pi_{a,b}\, |\Psi\ra=0$ implies that
any SA has a product form $|\Psi\ra=|\phi\ra_{ab}\otimes |\Psi'\ra$,
where $|\Psi'\ra$ is some state of the remaining $n-2$ qubits. Let us
show that $|\Psi\ra$ is a SA iff $|\Psi'\ra$ is a SA for a new
instance $y$ of quantum $2$-SAT defined on $n-2$ qubits. Indeed, for any
projectors $P,Q$ one has
\be
\label{reduction}
\left\{ \ba{rcl} P\, |\Psi\ra &=& |\Psi\ra, \\
Q\, |\Psi\ra &=& |\Psi\ra, \\
\ea \right. \quad \mbox{iff} \quad
 \left\{ \ba{rcl} P\, |\Psi\ra &=& |\Psi\ra, \\
P Q P\, |\Psi\ra &=& |\Psi\ra. \\
\ea\right.
\ee
Let us choose any pair of
qubits $c,d$ and set
\[
P=I-\Pi_{a,b} = |\phi\ra\la \phi|_{ab},
\quad
Q=I-\Pi_{c,d}.
\]
If the pairs $(a,b)$ and $(c,d)$ do not overlap, the right-hand side of
Eq.~(\ref{reduction}) is equivalent to
\[
|\Psi\ra=|\phi\ra_{ab}\otimes |\Psi'\ra, \quad \Pi_{c,d}\, |\Psi'\ra=0.
\]
On the other hand, if the pairs $(a,b)$ and $(c,d)$
overlap at one of the qubits, say $a=c$, then the right-hand side of
Eq.~(\ref{reduction}) is equivalent to
\[
|\Psi\ra=|\phi\ra_{ab}\otimes |\Psi'\ra, \quad
Q_d\, |\Psi'\ra=|\Psi'\ra,
\]
where $Q$ is a one-qubit self-adjoint operator implicitly defined by
\[
|\phi\ra\la\phi|_{ab} \,(I- \Pi_{a,d})\, |\phi\ra\la\phi|_{ab} = |\phi\ra\la\phi|_{ab}\otimes Q_d.
\]
Thus any $2$-qubit constraint imposed on $|\Psi\ra$
is equivalent to a $1$-qubit or $2$-qubit constraint imposed on $|\Psi'\ra$.
In other words, $|\Psi\ra$ is a SA for $x$ iff $|\Psi'\ra$ is a SA for a new
instance of quantum $2$-SAT $y$ defined on $n-2$ qubits.

\subsection{Projectors of rank two}
\label{subs:two}

Suppose that $r_{a,b}=2$ for some pair of qubits $a,b$.
We shall argue that qubits $a,b$ can be merged into a single logical qubit.
It will yield a new instance $y$ of quantum $2$-SAT defined on $n-1$ qubits equivalent to $x$.

Indeed, let $V\, : \, \CC^2\to \CC^2\otimes \CC^2$ be an isometry such that
\[
\Pi_{a,b}=I-(VV^\dag)_{ab}, \quad V^\dag V =I.
\]
The constraint $\Pi_{a,b}\, |\Psi\ra=0$ is equivalent to saying that
\[
|\Psi\ra=V_c\, |\Psi'\ra,
\]
where $|\Psi'\ra$ is some $(n-1)$-qubit state and the label $c$
refers to a qubit resulting from merging of $a$ and $b$.
The fact that $V$ is an isometry implies that for any pair of qubits
$f,g$
\be\label{isometry}
\Pi_{f,g}\, |\Psi\ra=0 \quad \mbox{iff} \quad
V_c^\dag \, \Pi_{f,g} \, V_c\, |\Psi'\ra=0.
\ee
If the pairs $(a,b)$ and $(f,g)$ do not overlap,
the right-hand side of Eq.~(\ref{isometry}) is equivalent to
$\Pi_{f,g}\, |\Psi'\ra=0$.
On the other hand, if the pairs $(a,b)$ and $(f,g)$ overlap at one of the
qubits, say $f=a$, then the right-hand side of Eq.~(\ref{isometry}) is equivalent to
\be
\label{new_constraint}
Q_{c,g}\, |\Psi'\ra = 0, \quad\mbox{where}\quad Q_{c,g}=V_c^\dag \, \Pi_{a,g}\, V_c.
\ee
Obviously, $Q_{c,g}$ is a self-adjoint operator that acts only on
two qubits $c$ and $g$.

Thus any $2$-qubit constraint imposed on $|\Psi\ra$
is equivalent to a $2$-qubit constraint imposed on $|\Psi'\ra$.
In other words, $|\Psi\ra$ is a SA for $x$ iff $|\Psi'\ra$ is a SA for a new
instance of quantum $2$-SAT $y$ defined on $n-1$ qubits.

\subsection{Homogeneous $2$-SAT}
From now on we can assume that $x$ is a homogeneous instance, i.e.,
 all non-zero projectors have rank one.
It will be convenient to describe a configuration of non-zero projectors
by a graph $G=(V,E)$, such that $V=\{1,\ldots,n\}$,
and $(a,b)\in E$ iff $\Pi_{a,b}\ne 0$.
Now we can reformulate quantum $2$-SAT as follows.

\noindent
{\bf Input:} A graph $G=(V,E)$ with $n$ vertices
and a list of rank two tensors
$\phi^{(a,b)}$ assigned to edges $(a,b)\in E$.

\noindent
{\bf Problem:} Decide whether there exists a non-zero
tensor $\psi$ of rank $n$
such that
\be\label{constraint}
\sum_{\alpha_a,\alpha_b}\, \phi^{(a,b)}_{\alpha_a,\alpha_b} \, \psi_{\alpha_1,\ldots,\alpha_n}=0 \quad \mbox{for any} \quad (a,b)\in E.
\ee

\noindent
{\it Comments:}
One should not confuse a rank of a tensor (number of its indexes) with
a rank of a projector discussed in previous subsections.
From now on we shall omit the summation symbol assuming
that all repeated indexes are contracted.
 Indexes of all tensors take values $0$ and $1$.
To avoid clutter in formulas, sometimes we shall identify tensors of rank two with $2\times 2$ matrices
and use matrix/vector multiplication.
Let us also agree that $\phi^{(a,b)}_{\alpha,\beta}=\phi^{(b,a)}_{\beta,\alpha}$.
We shall use a symbol $\epsilon$ for the fully antisymmetric tensor of rank two, i.e.,
\[
\epsilon=\left( \ba{cc} 0 & 1 \\ -1 & 0 \\ \ea \right).
\]

The following important observation provides a simple local rule for
generating new constraints.
\begin{lemma}
\label{lemma:key}
Let $\phi$, $\theta$ be arbitrary tensors of rank two
and $\psi$ be a tensor of rank three such that
\[
\phi_{\alpha,\beta}\, \psi_{\alpha,\beta,\gamma}=0
\quad
\mbox{and}
\quad
\theta_{\beta,\gamma}\, \psi_{\alpha,\beta,\gamma}=0.
\]
Then $\psi$ also obeys
\[
\omega_{\alpha,\gamma}\, \psi_{\alpha,\beta,\gamma} =0,
\quad
\mbox{where}
\quad
\omega_{\alpha,\gamma} = \phi_{\alpha,\beta}\, \epsilon_{\beta,\delta}\,
\theta_{\delta,\gamma}.
\]
\end{lemma}
Using matrix representation of
rank two tensors, one has simply $\omega=\phi\, \ep \, \theta$.

\noindent
{\bf Proof:}
 Indeed,
\begin{eqnarray}
\omega_{\alpha,\gamma}\, \psi_{\alpha,0,\gamma} &=&
\psi_{\alpha,0,\gamma} \left(
\phi_{\alpha,0}\, \theta_{1,\gamma} -
\phi_{\alpha,1} \theta_{0,\gamma}
\right)
\nn \\
&=&
-\psi_{\alpha,1,\gamma}\, \phi_{\alpha,1}\, \theta_{1,\gamma}
-
\psi_{\alpha,0,\gamma} \, \phi_{\alpha,1} \, \theta_{0,\gamma}
\nn \\
&=&
-\phi_{\alpha,1} \left(
\psi_{\alpha,1,\gamma}\, \theta_{1,\gamma} +
\psi_{\alpha,0,\gamma}\, \theta_{0,\gamma}
\right) =0. \nn
\end{eqnarray}
Analogously one proves that $\omega_{\alpha,\gamma}\, \psi_{\alpha,1,\gamma}=0$.
\begin{flushright}
$\Box$
\end{flushright}
A straightforward generalization of the lemma is this.
Suppose $\psi$ is a tensor of rank $n$ such that
\be
\label{eq:new1}
\phi^{(a,b)}_{\alpha_a,\alpha_b}\, \psi_{\alpha_1,\ldots,\alpha_n}=0
\quad
\mbox{and}
\quad
\phi^{(b,c)}_{\alpha_b,\alpha_c}\, \psi_{\alpha_1,\ldots,\alpha_n}=0
\ee
for some tensors $\phi^{(a,b)}$ and $\phi^{(b,c)}$ of rank two and some
integers $a\ne b\ne c$.
Then $\psi$ also
obeys a constraint
\be
\label{eq:new2}
\omega^{(a,c)}_{\alpha_a,\alpha_c}\, \psi_{\alpha_1,\ldots,\alpha_n}=0,
\ee
where
\be
\label{eq:new3}
\omega^{(a,c)}_{\alpha,\gamma}=\phi^{(a,b)}_{\alpha,\beta}\,
\epsilon_{\beta,\delta}\,\phi^{(b,c)}_{\delta,\gamma}.
\ee

Consider an instance $x$ of $2$-SAT specified by a graph $G=(V,E)$
and tensors $\phi^{(a,b)}$, $(a,b)\in E$.
We can try to use Lemma~\ref{lemma:key}
and its corollary Eqs.~(\ref{eq:new2},\ref{eq:new3})
to generate new constraints on $\psi$ from the existing ones.
Indeed, consider any pair of edges $(a,b)\in E$ and $(b,c)\in E$.
Let $\omega^{(a,c)}$ be a tensor defined in Eq.~(\ref{eq:new3}).
One  has to consider the following possibilities:
\begin{enumerate}
\item $\omega^{(a,c)}=0$. We get no additional constraints on $\psi$.

\item $\omega^{(a,c)}\ne 0$, but $(a,c)\in E$ and $\omega^{(a,c)}$
is proportional to $\phi^{(a,c)}$. Again, we get no additional constraints on $\psi$.

\item $\omega^{(a,c)}\ne 0$ and $(a,c)\notin E$.
In this case Eq.~(\ref{eq:new2}) provides a new constraint on $\psi$.
Let us add $(a,c)$ into the list of edges $E$ and set
$\phi^{(a,c)}=\omega^{(a,c)}$.

\item $\omega^{(a,c)}\ne 0$, $(a,c)\in E$, and $\omega^{(a,c)}$
is not proportional to $\phi^{(a,c)}$.
In this case we get two independent constraints for the pair of qubits
$a$ and $c$. They leave us with a two-dimensional forbidden  subspace
for this pair of qubits. Thus we can apply ideas of the previous
subsection to merge $a$ and $c$ into a single logical qubit
and obtain a new instance of quantum $2$-SAT with $n-1$ qubits which is equivalent
to the original one.
\end{enumerate}

Let us keep trying to generate new constraints by probing
different pairs of edges until we either
encounter the case~4
reducing the number of
qubits from $n$ to $n-1$ (the algorithm terminates with Output~3), or after $O(n^3)$
steps we arrive to a
homogeneous instance with
{\it a complete set of constraints}  which is defined below.

\begin{definition}
A graph $G=(V,E)$ and a family of rank two tensors
$\phi^{(a,b)}$, $(a,b)\in E$ constitute  a complete set of constraints
iff for any pair of edges $(a,b)\in E$ and $(b,c)\in E$
one of the following is true
\begin{itemize}
\item If $(a,c)\in E$ then $\phi^{(a,b)}\, \epsilon\, \phi^{(b,c)}$ is
either proportional to $\phi^{(a,c)}$ or zero,
\item If $(a,c)\notin E$ then $\phi^{(a,b)}\, \epsilon\, \phi^{(b,c)}=0$.
\end{itemize}
\end{definition}

This definition just says that any attempt to generate a new constraint on $\psi$
using Lemma~\ref{lemma:key} would fail.
\begin{lemma}
Any homogeneous instance of quantum $2$-SAT with a complete set of
constraints has a satisfying assignment.
It can be chosen as a product of one-qubit state.
\end{lemma}
{\it Remark:} A proof given below is constructive. It allows one to find
a product satisfying assignment in a time $O(n)$, $n=|V|$.\\
{\bf Proof:}
Let $\{ \phi^{(a,b)} \}$, $(a,b)\in E$ be a family of rank two tensors constituting
a complete set of constraints.
We shall construct a satisfying assignment $\psi$ which is a product of tensors of rank one:
\[
\psi_{\alpha_1,\alpha_2,\ldots,\alpha_n}=\psi^{(1)}_{\alpha_1}\,
\psi^{(2)}_{\alpha_2}\cdots \psi^{(n)}_{\alpha_n}.
\]
Let us start from choosing an arbitrary $\psi^{(1)}$.
Denote $\nrb{a}\subset V$ a set of all nearest neighbors of a vertex $a$.
The next step is to assign states to all vertices $a\in \nrb{1}$
according to $\psi^{(a)}=\epsilon\, (\phi^{(1,a)})^T\, \psi^{(1)}$,
or in tensor notations,
\be
\label{eq:greedy}
\psi^{(a)}_\alpha= \epsilon_{\alpha,\beta}\, \phi^{(1,a)}_{\gamma,\beta}
\, \psi^{(1)}_\gamma, \quad a\in \nrb{1}.
\ee
Let us verify that all constraints on edges incident to the vertex $1$ are satisfied.
Indeed,
\[
\phi^{(1,a)}_{\delta,\alpha} \, \psi^{(1)}_\delta\, \psi^{(a)}_\alpha =
\phi^{(1,a)}_{\delta,\alpha}\, \psi^{(1)}_\delta \,
\epsilon_{\alpha,\beta}\, \phi^{(1,a)}_{\gamma,\beta}\, \psi^{(1)}_\gamma=0
\]
because $\ep$ is an antisymmetric tensor. This is equivalent to
the desired constraints
\[
\phi^{(1,a)}_{\alpha_1,\alpha_a}\, \psi_{\alpha_1,\ldots,\alpha_n}=0
\quad \mbox{for any} \quad a\in\nrb{1}.
\]
The next step is to verify that all constraints on  edges connecting
two vertices from $\nrb{1}$ are automatically satisfied.
Indeed consider any pair of edges $(1,a)\in E$ and $(1,c)\in E$ such that
$(a,c)\in E$ as well. Then by definition of a complete set of constraints
we have $\phi^{(a,c)}=\phi^{(a,1)}\, \epsilon\, \phi^{(1,c)}$
(up to some overall factor).
Since we have already fulfilled the constraints on the edges
$(a,1)$ and $(1,c)$, we can apply Lemma~\ref{lemma:key} (with $b=1$)
to infer that
\[
\phi^{(a,c)}_{\alpha,\gamma} \, \psi^{(a)}_\alpha \, \psi^{(c)}_\gamma=0.
\]
This is equivalent to the desired constraints
\[
\phi^{(a,c)}_{\alpha_a,\alpha_c}\, \psi_{\alpha_1,\ldots,\alpha_n}=0
\quad \mbox{for any} \quad a,c\in\nrb{1},
\quad (a,c)\in E.
\]
Let us split the set of all vertices $V$ into two subsets,
$V=V_{close}\cup V_{far}$, where $V_{close}$ consists of the vertex $1$
and its nearest neighbors, while $V_{far}$ consists of all other vertices
(those having distance two or greater from the vertex $1$).
All vertices in $V_{close}$ have been already assigned a state.
This assignment satisfies the constraints on all edges
having both ends in $V_{close}$.
We shall now prove that this assignment also satisfies
the constraints on any edge having one end in $V_{close}$
and another end in $V_{far}$, for {\it any} choice of assignments
in $V_{far}$.

Indeed,
consider any edge $(b,c)\in E$, such that $b\in\nrb{1}$
and $c\in V_{far}$. Since
$(1,c)\notin E$,
by definition of a complete set of constraints we have
\be
\label{eq:pep=0}
\phi^{(1,b)}\, \epsilon\, \phi^{(b,c)} =0.
\ee
Let us verify that the constraint on the edge $(b,c)$ is
fulfilled
for {\it all } choices of an assignment
$\psi^{(c)}$, i.e.,
\be
\label{eq:verify1}
\phi^{(b,c)}_{\beta,\gamma} \, \psi^{(b)}_\beta\, \psi^{(c)}_\gamma=0
\quad \mbox{for all}
\quad
\psi^{(c)}.
\ee
It suffices to verify that
\be
\label{eq:verify2}
\phi^{(b,c)}_{\beta,\gamma}\, \psi^{(b)}_\beta=0.
\ee
Indeed, from Eq.~(\ref{eq:greedy}) we have
\[
\psi^{(b)}_\beta = \epsilon_{\beta,\alpha} \,
\phi^{(1,b)}_{\delta,\alpha}\, \psi^{(1)}_\delta.
\]
Therefore
\[
\phi^{(b,c)}_{\beta,\gamma}\, \psi^{(b)}_\beta =
\phi^{(b,c)}_{\beta,\gamma} \,
\epsilon_{\beta,\alpha}\,
\phi^{(1,b)}_{\delta,\alpha}\,
\psi^{(1)}_\delta =
-( \phi^{(1,b)}\, \epsilon \, \phi^{(b,c)})_{\delta,\gamma}\, \psi^{(1)}_\delta =0,
\]
see Eq.~(\ref{eq:pep=0}). Thus we have proved Eq.~(\ref{eq:verify2}),
and therefore Eq.~(\ref{eq:verify1}). This is equivalent to the desired
constraints
\be
\label{eq:auto}
\phi^{(b,c)}_{\alpha_b,\alpha_c}\, \psi_{\alpha_1,\ldots,\alpha_n}=0
\quad \mbox{for any} \quad b\in \nrb{1}, \quad c\in V_{far}.
\ee
for any choice of an assignment at the vertices $c\in V_{far}$.

The only constraints that we have not verified yet are those
sitting on edges having both ends in $V_{far}$.
One can easily check that a
family of rank two tensors $\{ \phi^{(a,b)}\}$,
$a,b\in V_{far}$, $(a,b)\in E$, constitutes a complete set
of constraints (since its definition is local).
Thus we can keep applying the same algorithm to
assign states to vertices from $V_{far}$.
In this way we will finally end up with a SA for all $n$ qubits.

\begin{flushright}
$\Box$
\end{flushright}

\section{Quantum $4$-SAT is \QMAe{} complete }
\label{sec:4sat}

Let us start from defining a complexity class \QMAe.
Throughout this section all quantum circuits are
assumed to have a data input register $R_{in}$,
a witness input register $R_{wit}$, and an output
register $R_{out}$.
We shall assume that all qubits of the data input
register $R_{in}$ are set initially in $|0\ra$ state.
Final measurement involves measurement of each qubit from $R_{out}$
in the standard $\{|0\ra,|1\ra\}$ basis.
For any quantum circuit $U$ and witness input state
$|\psi_{wit}\ra$ define an {\it acceptance probability}
$AP(U,\psi_{wit})$ as a
probability for the final measurement to yield the outcome $0$
in every qubit of $R_{out}$.

Let $\calF\subset \CC$ be a field that we use for exact representation of
complex numbers (see the remark in Introduction).
We shall consider a gate set $\calG$ consisting of all
three-qubit unitary operators whose matrix
elements in the standard basis belong to $\calF$
(we include three-qubit gates into $\calG$ just to simplify the proofs).

\begin{definition}
A language $L=L_{yes}\cup L_{no}\subset \Sigma^*$ belongs to \QMAe{} iff there
exists a polynomial $p$ and
uniform family of quantum circuits $\{U(x)\}$, $x\in L$,
such that $U(x)$ has a length at most $p(|x|)$,
$U(x)$ uses only gates from $\calG$,
$|R_{in}|+|R_{wit}|+|R_{out}|\le p(|x|)$,
and
\begin{itemize}
\item If $x\in L_{yes}$ then there exists a witness state $|\psi_{wit}\ra$
such that $AP(U(x),\psi_{wit})=1$;
\item If $x\in L_{no}$ then for any witness state $|\psi_{wit}\ra$
one has $AP(U(x),\psi_{wit})<1-1/p(|x|)$.
\end{itemize}
\label{def:QMAe}
\end{definition}

Since there is a polynomial  gap between acceptance probabilities corresponding to
positive and negative instances, we conclude that
\QMAe~$\subseteq$~\QMA.
From now on we shall assume that for any instance of quantum $k$-SAT
projectors $\Pi_S$ have matrix elements from the field $\calF$.

\begin{lemma}
Quantum $k$-SAT belongs to \QMAe{} for any constant $k$.
\end{lemma}
{\bf Proof:}
Let $\{\Pi_S\}$, $S\subset \{1,2,\ldots,n\}$, $|S|=k$ be an instance of
quantum $k$-SAT.
We shall construct a quantum circuit that measures
eigenvalues of $\Pi_S$ and writes the outcomes to output register $R_{out}$.
More strictly, for each subset $S$, $|S|=k$, introduce an auxiliary
qubit $q(S)$ initially prepared in the $|0\ra$ state, and consider a unitary operator
\[
U_S= \Pi_S\otimes \sigma^x_{q(S)} + (I-\Pi_S)\otimes I_{q(S)}.
\]
It acts on a witness input register $R_{wit}$ of size $n$ and the
auxiliary qubit $q(S)$. Note that $U_S$ acts non-trivially only on a
subset of $k+1$ qubits.
According to Lemma~\ref{lemma:app}, Appendix~A, the operator
$U_S$ can be exactly represented by a quantum circuit
of a size $poly(2^k)$ with gates from the set $\calG$.

For any input witness state $|\psi_{wit}\ra$ one has
\be
\label{eq:US}
U_S\, |\psi_{wit}\ra\otimes |0\ra = \Pi_S\, |\psi_{wit}\ra\otimes |1\ra +
(I-\Pi_S)\,|\psi_{wit}\ra \otimes |0\ra.
\ee
Denote $p_S$ a probability to measure the qubit $q(S)$ in the state $|1\ra$
after the application of $U_S$.
From Eq.~(\ref{eq:US}) one can easily get
\[
p_S=\la \psi_{wit}|\Pi_S|\psi_{wit}\ra.
\]

Let us define  the output register $R_{out}$ as a collection of all ancillary qubits $q(S)$,
one qubit for each subset $S$.
Choose some order on the set of subsets $S$ and
define a verifying circuit as
$U=\prod_S U_S$.
By construction, $U$ accepts on a witness state $|\psi_{wit}\ra$ iff
all the qubits $q(S)$ have been measured in $|0\ra$.
A probability of that event can be bounded as
\be
\label{eq:bounds}
1-\sum_S p_S \le AP(U,\psi_{wit}) \le  1-\max_S p_S.
\ee
If there exists a satisfying assignment, one can choose
$|\psi_{wit}\ra$ such that $p_S=0$ for all $S$, and thus
$AP(U,\psi_{wit})=1$.
On the other hand, if there is no satisfying assignment,
we are promised that for any witness state
$\sum_S p_S \ge \ep$, where $\ep=1/n^\alpha$, $\alpha=O(1)$,
see Introduction.
Therefore
\[
\max_S p_S\ge \frac{\ep}{{n \choose k}} \ge n^{-k}\, \ep.
\]
Combining it with Eq.~(\ref{eq:bounds}) we get
$AP(U,\psi_{wit})\le 1- 1/n^{k+\alpha}$ for any witness state $|\psi_{wit}\ra$.
Now one can easily choose
a proper polynomial $p$ to meet all the requirements of Definition~\ref{def:QMAe}.
\begin{flushright}
$\Box$
\end{flushright}

\begin{lemma}
\label{lemma:complete}
Quantum $k$-SAT is \QMAe-complete for any constant $k\ge 4$.
\end{lemma}

{\bf Proof of Lemma~\ref{lemma:complete}:}
It suffices to prove that quantum $4$-SAT is \QMAe-complete.
Let
$U=U_L\cdots U_2\, U_1$, $U_j\in \calG$,
be a quantum circuit  operating on $N$ qubits
with input data and witness registers $R_{in}$ and $R_{wit}$,
and output register $R_{out}$.
Denote $|\psi_{in}\ra=|0,\ldots,0\ra$ and $|\psi_{wit}\ra$ the initial states
of $R_{in}$ and $R_{wit}$.
Without loss of generality we can assume that
$N=|R_{in}| + |R_{wit}|$.
The collection of $N$ qubits $R_{in}\cup R_{wit}$ there the computation goes
will be referred to as a {\it computational register}. It is described by
a Hilbert space
\[
\calH_{\comp}=(\CC^2)^{\otimes N}.
\]
Let us define $L+1$ intermediate computational states
\[
|Q_0\ra=|\psi_{in}\ra\otimes |\psi_{wit}\ra\quad \mbox{and} \quad |Q_{t}\ra=U_t\, |Q_{t-1}\ra,
\quad t=1,\ldots,L.
\]
Introduce an auxiliary Hilbert space
\[
\calH_{\clock} = (\CC^4)^{\otimes L}
\]
that describes a {\it clock register}
composed of  $L$ four-dimensional {\it clock particles},
one particle associated with each gate in the circuit.
Each clock particle can be represented by a pair of qubits,
so if we want to keep all projectors $4$-local, we can afford
only projectors acting on $n_c$ clock particles and $n_q$ computational
qubits, where
\[
(n_c,n_q)\in \{ (2,0), \, (1,2), \, (0,4) \}.
\]
Later on we shall define $2L$ orthonormal {\it legal clock states}
\[
|C_1\ra,\, |C_1'\ra,\ldots, |C_L\ra,\, |C_L'\ra \in \calH_{\clock}.
\]
Our first goal is to design a Hamiltonian $H(U)$
such that
\begin{itemize}
\item $H(U)$ acts on a Hilbert space $\calH=\calH_{\clock}\otimes \calH_{\comp}$,
\item $H(U)$ is a sum of $4$-qubit projectors;
\item The zero-subspace of $H(U)$ is spanned by vectors
\be
\label{eq:Omega}
|\Omega(\psi_{wit})\ra= \sum_{t=1}^L
|C_t\ra\otimes |Q_{t-1}\ra  + |C_t'\ra\otimes |Q_t\ra, \quad |Q_0\ra = |\psi_{in}\ra\otimes |\psi_{wit}\ra.
\ee
\end{itemize}
The state $|\Omega(\psi_{wit})\ra$ represents the whole history of the computation
starting from a witness state $|\psi_{wit}\ra$.
Note that
 the gate $U_t$ is applied when the clock register
goes from the state $|C_t\ra$ to the state $|C_t'\ra$.
On the other hand, as the clock register goes from
$|C_t'\ra$ to $|C_{t+1}\ra$, the state of the computational
register does not change.

Now let us specify the legal clock states.
Basis states of a single clock particle will be
denoted as $|u\ra$ --- {\it unborn}, $|d\ra$ --- {\it dead},
$|a1\ra$ --- {\it active phase~1}, $|a2\ra$ --- {\it active phase~2}
(this terminology is partially borrowed from~\cite{ADKLLR}.
As the time goes forward, each clock particle evolves from
the unborn phase to active phase~1, then to active phase~2 and finally ends up
in the dead phase. Below we list the eight legal clock states for $L=4$
as an example.
\be
\label{eq:example}
\ba{rcl}
|C_1\ra&=&|a1,u,u,u\ra,  \\
|C_1'\ra&=&|a2,u,u,u\ra,  \\
|C_2\ra&=&|d,a1,u,u\ra,  \\
|C_2'\ra&=&|d,a2,u,u\ra,  \\
\ea
\quad
\ba{rcl}
|C_3\ra&=&|d,d,a1,u\ra,  \\
|C_3'\ra&=&|d,d,a2,u\ra,  \\
|C_4\ra&=&|d,d,d,a1\ra,  \\
|C_4'\ra&=&|d,d,d,a2\ra.  \\
\ea
\ee
In general, legal clock states are defined
as basis vectors of $\calH_{\clock}$ that obey the following
constraints:
\begin{enumerate}
\item The particle $1$ is either active or dead,
\item The particle $L$ is either active or unborn,
\item There is at most one active particle,
\item If a particle $j$ is dead for all $1\le k<j$
a particle $k$ is dead,
\item If a particle $j$ is unborn then for all $j<k\le L$
a particle $k$ is unborn,
\item If a particle $j$ is dead then a particle $j+1$
is either dead or active.
\end{enumerate}
One can easily check that there are only $2L$ basis states
satisfying these rules. Each of these states
has exactly one active particle
(in either active phase~1 or active phase~2) at some position
$1\le j\le L$, dead particles at positions $k<j$ and
unborn particles at positions $k>j$,
\[
|C_j\ra=|\underbrace{d,\ldots,d}_{j-1},a1,\underbrace{u,\ldots,u}_{L-j}\ra,
\quad
|C_j'\ra=|\underbrace{d,\ldots,d}_{j-1},a2,\underbrace{u,\ldots,u}_{L-j}\ra.
\]
Denote
\[
\calH_{\legal}=\mbox{Linear Span}(|C_1\ra,|C_1'\ra,\ldots,|C_L\ra,|C_L'\ra)\subset
\calH_{\clock}
\]
a subspace spanned by the legal clock states.
Let us define a Hamiltonian $H_{\clock}$ acting on the
space $\calH_{\clock}$ as
\[
H_{\clock}= \sum_{j=1}^6 H_{\clock}^{(j)},
\]
\begin{eqnarray}
H_{\clock}^{(1)}&=& |u\ra\la u|_1, \nn \\
H_{\clock}^{(2)} &=& |d\ra\la d|_L, \nn \\
H_{\clock}^{(3)}&=&\sum_{1\le j<k\le L} ( |a1\ra\la a1| + |a2\ra\la a2|)_j \otimes (|a1\ra\la a1| + |a2\ra\la a2|)_k,
\nn \\
H_{\clock}^{(4)}&=&\sum_{1\le j<k\le L} ( |a1\ra\la a1| + |a2\ra\la a2| + |u\ra\la u|)_j \otimes |d\ra\la d|_k,
\nn \\
H_{\clock}^{(5)}&=&\sum_{1\le j<k\le L} |u\ra\la u|_j \otimes
 ( |a1\ra\la a1| + |a2\ra\la a2| + |d\ra\la d|)_k,
\nn \\
H_{\clock}^{(6)}&=&\sum_{1\le j\le L-1} |d\ra\la d|_j \otimes |u\ra\la u|_{j+1}.\nn
\end{eqnarray}
The six terms of $H_{\clock}$ correspond to the six constraints listed above,
so that
\be
|\psi\ra\in \calH_{\legal} \quad \mbox{iff} \quad
H_{\clock}\, |\psi\ra =0.
\ee
Note that $H_{\clock}$ is a sum of $4$-qubit projectors.
Denote $\Pi$ an orthogonal projector onto a subspace $\calH_{\legal}\otimes\calH_{\comp}$,
\[
\Pi=\sum_{j=1}^L (|C_j\ra\la C_j| + |C_j'\ra\la C_j'|)\otimes I_{\comp}.
\]

Define a Hamiltonian $H_{\init}$ as
\be
\label{eq:init}
H_{\init} =  |a1\ra\la a1|_1\otimes \left( \sum_{b\in R_{in}} |1\ra\la 1|_b \right).
\ee
The only legal clock state having $|a1\ra$ in the first position is $|C_1\ra$.
Thus $H_{\init}$ penalizes any qubit of the input data
register $R_{in}$ for being in a state
$|1\ra$ provided that the clock register's state is $|C_1\ra$.
A restriction of $H_{\init}$ onto the subspace $\calH_{\legal}\otimes \calH_{\comp}$
is therefore
\be
\label{eq:piece0}
\Pi \, H_{\init} \, \Pi = |C_1\ra\la C_1|\otimes \left( \sum_{b\in R_{in}} |1\ra\la 1|_b \right).
\ee
Note that $H_{\init}$ is a sum of $3$-qubit projectors.

Define a Hamiltonian $H_{\prop}$ as
\be
\label{eq:prop1}
H_{\prop}=\sum_{t=1}^L H_{\prop,t} + \sum_{t=1}^{L-1} H_{\prop,t}',
\ee
where
\be
 \label{eq:prop2}
H_{\prop,t}= \frac12 \left[ \left( |a1\ra\la a1| + |a2\ra\la a2|
\right)_t \otimes I_{\comp} - |a2\ra\la a1|_t \otimes U_t -
|a1\ra\la a2|_t \otimes U_t^\dag \right],
\ee
and
\be
\label{eq:prop3}
 H_{\prop,t}'= \frac12 \left( |a2,u\ra\la a2,u| +
|d,a1\ra\la d,a1| - |d,a1\ra\la a2,u| - |a2,u\ra\la d,a1|
\right)_{t,t+1}\otimes I_{\comp}.
\ee
The operators $H_{\prop,t}$
and $H_{\prop,t}'$ are $4$-qubit  projectors. By
inspecting the example Eq.~(\ref{eq:example}) one can easily check
that
\begin{itemize}
\item The only legal clock state having $a1$
at position $t$ is $|C_t\ra$;
\item The only legal clock state having $a2$
at position $t$ is $|C_t'\ra$;
\item  The only legal clock state
having $(d,a1)$ at positions $t,t+1$ is $|C_{t+1}\ra$;
\item
The only legal clock state having $(a2,u)$ at positions $t,t+1$
is $|C_t'\ra$.
\end{itemize}
Thus a restriction of $H_{prop,t}$ onto the
subspace $\calH_{legal}\otimes \calH_{\comp}$ is
\be
\label{eq:piece1}
\Pi\, H_{prop,t}\, \Pi =\frac12
\left[
\left(
|C_t\ra\la C_t| + |C_t'\ra\la C_t'|
\right)\otimes I_{\comp} -
|C_t'\ra\la C_t|\otimes U_t - |C_t\ra\la C_t'|\otimes U_t^\dag
\right].
\ee
Analogously, a restriction of $H_{prop,t}'$ onto the subspace
$\calH_{legal}\otimes \calH_{\comp}$ is
\be
\label{eq:piece2}
\Pi\, H_{prop,t}'\, \Pi=\frac12
\left(
|C_t'\ra\la C_t'| + |C_{t+1}\ra\la C_{t+1}|
-
|C_{t+1}\ra\la C_t'|  - |C_t'\ra\la C_{t+1}|
\right)\otimes I_{\comp}.
\ee
Define a Hamiltonian
\[
H(U)=H_{\init} + H_{\clock} + H_{\prop}.
\]
Combining Eqs.~(\ref{eq:piece0},\ref{eq:piece1},\ref{eq:piece2})
one can easily check  that the zero-subspace of $H(U)$ is indeed
spanned by computational history states $|\Omega(\psi_{wit})\ra$.

One remains to introduce an extra term into $H(U)$ that
is responsible for the final measurement of $R_{out}$.
Define
\[
H_{\out}=|a2\ra\la a2|_L\otimes \left(\sum_{b\in R_{out}} |1\ra\la 1|_b\right).
\]
The only legal clock state having $a2$ at the position $L$ is $|C_L'\ra$.
Thus a restriction of $H_{\out}$ onto the subspace $\calH_{\legal}\otimes \calH_{\comp}$
is
\[
\Pi\, H_{\out}\, \Pi = |C_L'\ra\la C_L'|\otimes \left(\sum_{b\in R_{out}} |1\ra\la 1|_b\right).
\]
Therefore $H_{\out}$ penalizes any qubit
of the output register for being in the state $|1\ra$
provided that the clock register is in the state $|C_L'\ra$.
Besides, $H_{\out}$ is a sum of $3$-qubit projectors.

We summarize that a Hamiltonian
\be
\label{eq:H}
H=H_{\init} + H_{\clock} + H_{\prop} + H_{out}
\ee
has zero ground state energy iff
there exists input witness state $|\psi_{wit}\ra$ such that
$AP(U,\psi_{wit})=1$.
By construction, $H$
operates on $N+2L$ qubits, and it can be represented as
a sum of $4$-qubit projectors:
\be
\label{eq:H'}
H=\sum_S \Pi_S, \quad S\subseteq \{1,\ldots,N+2L\}, \quad |S|=4.
\ee
Thus
the property ``$H$ has zero ground state energy''
is equivalent to the quantum $4$-SAT $\{\Pi_S\}$
having a satisfying assignment.

Let $L=L_{yes}\cup L_{no}$ be a language from \QMAe, $x\in L$ be a binary string, and
$U(x)$ be a verifying circuit
see Definition~\ref{def:QMAe}.
Using the majority voting to amplify the gap in acceptance probabilities,
see~\cite{KSV}, we can assume that
\begin{itemize}
\item If $x\in L_{yes}$ then $AP(U,\psi_{wit})=1$ for some input witness state $|\psi_{wit}\ra$,
\item If $x\in L_{no}$ then $AP(U,\psi_{wit})\le \ep$, $\ep=1/p(|x|)$, for all $|\psi_{wit}\ra$,
\end{itemize}
where $U$ is a circuit implementing several copies of $U(x)$ and the majority
voting (obviously it can be realized using the gate set $\calG$).
The polynomial $p$ above may be different from the one in Definition~\ref{def:QMAe}.

To complete the proof of the lemma we have to show that
for negative instances,
$x\in L_{no}$, the ground state energy of the  Hamiltonian Eq.~(\ref{eq:H})
is not too small, i.e., $\la \Psi|H|\Psi\ra\ge 1/q(|x|)$ for any
$|\Psi\ra$, where $q$ is some polynomial.
It can be done using ideas from~\cite{KSV}.
Indeed, a decomposition
\[
\calH=\calH_{\legal}\otimes \calH_{\comp} \oplus \calH_{\legal}^\perp \otimes
\calH_{\comp}
\]
is invariant under $H$. Let $H'$ be a restriction of $H$ onto $\calH_{\legal}\otimes \calH_{\comp}$.
One can easily check that $H'$ is exactly the Hamiltonian that one would
assign to a quantum circuit $\tilde{U}$ using the construction of~\cite{KSV},
where $\tilde{U}$ is obtained from $U$ by appending the identity gate
to each gate of $U$ (if $U$ has a length $L$ then $\tilde{U}$ has a length $2L$).
It was shown in~\cite{KSV} that $H'$ has the ground state energy
at least $c(1-\sqrt{\ep})(2L)^{-3}$, where $c$ is a constant and $\ep$ is
defined above.
It may be only polynomially small in $|x|$.

Let $H''$ be a restriction of $H$ onto $\calH_{\legal}^{\perp}\otimes \calH_{\comp}$.
Since any state from $\calH_{\legal}^{\perp}$ violates at least one
constraint from $H_{\clock}$, the ground state energy of $H''$
is at least $1$.

\begin{flushright}
$\Box$
\end{flushright}

{\it Remark~1:}
One could also try to consider clock particles with only three
states: unborn, active, and dead. It is possible to design
a proper dynamics of the clock register, such that the
computational history state can be specified by $3$-local
projectors~\cite{DV}. Each projector involves a triple of particles
with dimensions $3\times 2\times 2$.

{\it Remark~2:} If one modifies Definition~\ref{def:QMAe}
such that
the acceptance probability corresponding to
negative instances is $1-\ep$, where $\ep>0$ may
be arbitrarily small, the corresponding class would
be hard for the polynomial hierarchy, as it would
contain the exact counting class \CP, see~\cite{Fenner}.
The same remark concerns quantum $k$-SAT, $k\ge 4$,
without the polynomial gap promise.

\begin{acknowledgments}
The author would like to thank David DiVincenzo,
Roberto Oliveira, and Barbara  Terhal
for numerous discussions and useful comments.
Inspiring correspondence on
quantum complexity classes
with Mikhail Vyalyi is acknowledged.
Part of this work was carried out when the author was a
member of the Institute for Quantum Information, Caltech,
supported by the National Science Foundation under
grant number EIA-0086038.
\end{acknowledgments}

\appendix
\section{Quantum circuits over a fixed field}

\begin{lemma}
\label{lemma:app}
Let $V$ be a unitary operator acting on $k$ qubits. Suppose that
matrix elements of $V$  in
the standard basis belong to a field $\calF\subset \CC$.
Then $V$ can be exactly represented
by a quantum circuit of size $poly(k) \cdot 2^{2k}$ with three-qubit gates
whose matrix elements in the standard basis also belong to the field $\calF$.
\end{lemma}
{\bf Proof:}
The first step
directly follows the proof of universality
of two-qubit gates, see~\cite{Barenco}.
Let us decompose $V$ as
$V=\tilde{W}_1 \tilde{W}_2 \cdots \tilde{W}_L$, where each operator $\tilde{W}_j$
is a direct sum of 2x2 unitary on a two-dimensional subspace
spanned by some pair of basis vectors and the identity operator
on a subspace spanned by the remaining $2^k-2$ basis vectors.
Accordingly, $L=2^{k-1}(2^k-1)$ is the number of such subspaces.
Matrix elements of the operators $\tilde{W}_j$ belong to the same
field $\calF$.
Using swaps of qubits and $\sigma^x$ gates one can transform
any operator $\tilde{W}_j$ as above into a controlled one-qubit gate
with one target qubit and $k-1$ control qubits,
$\tilde{W}_j\equiv \Lambda_{k-1}(U_j)$.  Here $U_j$ is
a one-qubit gate with matrix elements from $\calF$.

Consider a classical unitary operator $\Omega$ that reversibly computes
logical $AND$ of $k$ bits $x_1,\ldots,x_{k-1}$, i.e.,
\[
\Omega\, |a\ra\otimes |x_1,x_2,\ldots,x_{k-1}\ra =
|a\oplus AND(x_1,\ldots,x_{k-1})\ra\otimes |x_1,x_2,\ldots,x_{k-1}\ra.
\]
Here $|a\ra$ refers to an ancillary qubit.
Since three-bit classical gates constitute universal basis for classical reversible
computation, we can implement $\Omega$ (probably using additional $|0\ra$ ancillas)
by a circuit of size $poly(k)$ with three-qubit gates whose matrix elements are only $0$
and $1$.
Now one can implement $\Lambda_{k-1}(U_j)$ as follows (we label the ancillary qubit
by $A$ and the target qubit by $T$):\\
(1) Set $A$ to $|0\ra$,\\
(2) Apply $\Omega$ to $A$ and the $k-1$ control qubits,\\
(3) Apply $\Lambda(U_j)$ to $A$ and $T$ such that $A$ is the control qubit,\\
(4) Apply $\Omega^{-1}$ to $A$ and the $k-1$ control qubits.

One can easily check that the ancillary qubit ends up in the $|0\ra$ state,
while the $k-1$ control qubits and the target qubit are acted on by $\Lambda_{k-1}(U_j)$.
By composing these circuits for each $j$ one gets a circuit representing
$V$.
\begin{flushright}
$\Box$
\end{flushright}

\section{Efficient algorithm for classical $2$-SAT}
\label{sec:classical}

Let $x=(x_1,\ldots,x_n)$ be a binary string. A Boolean function
$l(x)$ is called a {\it literal} if $l(x)=x_a$ or $l(x)=(\neg x_a)$
for some $1\le  a\le n$. Given $m$ Boolean functions (clauses)
$C_j(x)=l_j(x)\vee l_j'(x)$, where $l_j$ and $l_j'$ are literals, a
classical $2$-SAT problem is to determine whether there exists a
string $x$ such that $C_j(x)=1$ for all $j=1,\ldots,m$. Below we
briefly describe a well-known algorithm for a classical $2$-SAT that
runs in a time $O(n+m)$.

Without loss of generality $l_j'\ne (\neg l_j)$
(otherwise $C_j(x)=1$ for all $x$).
Consider a directed graph $G=(V,E)$, whose vertices
are literals, i.e.,
\[
V=\{x_1,\ldots,x_n,\bar{x}_1,\ldots,\bar{x}_n\},
\]
and whose edges are pairs of literals $(l,l')$ that appear in the
same clause as shown below:
\[
E=\{ (l,l')\, : \,  C_j=l\vee (\neg l') \quad \mbox{for some} \quad j\}.
\]
Each clause $C_j$ contributes to one or two edges of $G$.
Let us split $V$ into strongly connected components (SCC). (By definition,
$l$ and $l'$ belong to the same SCC iff there exists a path
from $l$ to $l'$ and a path from $l'$ to $l$.) Obviously, one can identify all
SCCs of $G$ in a time
$O(n+m)$.
\begin{lemma}
A satisfying assignment $x$ exists iff for any $l\in V$
vertices $l$ and $\neg l$ belong to different SCCs.
\end{lemma}
{\bf Proof:}

\noindent
(a) Suppose for some $l,l'\in V$ there exists a path from $l$ to $l'$.
If $x$ is a SA then $l(x)=0$ implies $l'(x)=0$. If $l$ and $\neg l$ belong to
the same SCC, then $l(x)=0$ implies $\neg l(x)=0$ and $\neg l(x)=0$ implies $l(x)=0$.
This is a contradiction.

\noindent
(b) Suppose that $l$ and $\neg l$ belong to different SCCs for all $l\in V$.
Let us find a SA.
Consider a directed graph $\tilde{G}=(\tilde{V},\tilde{E})$ whose vertices
are SCCs of $G$ and whose edges are pairs $(S,S')$ of SCCs of $G$
such that there exists a path from $S$ to $S'$.
By definition, $\tilde{G}$ is acyclic. Using the topological sorting algorithm
one can order vertices of $\tilde{G}$ such that $(S,S')\in \tilde{E}$
implies $S<S'$. It can be done in a time $O(n+m)$.

By definition of $G$, for any vertices $l,l'\in V$ a path from $l$ to $l'$ and a path from
$\neg l'$ to $\neg l$ exist or do not exist simultaneously.
Thus for any vertex $S$ of $\tilde{G}$ there exists a unique vertex $\neg S$
of $\tilde{G}$ such that $\neg S=\{ l\in V \, : \, \neg l\in S\}$.

For all vertices $S$ of $\tilde{G}$ do the following:
If $S<(\neg S)$ then
set $l=1$ for all $l\in S$ and set $l=0$ for all $l\in (\neg S)$.
If $S>(\neg S)$ then set $l=0$ for all $l\in S$ and
set $l=1$ for all $l\in (\neg S)$.
We end up with some assignment $x$. Let us check that $C_j(x)=1$
for all $j$. Indeed, otherwise we would have
$l_j(x)=l_j'(x)=0$ for some $j$. Let $S$ and $S'$ be SCCs of
$l_j$ and $l_j'$. We have $S>\neg S$ and $S'>\neg S'$.
On the other hand, $S<\neg S'$ and $S'<\neg S$. This is a contradiction.
Thus $x$ is a SA.

\begin{flushright}
$\Box$
\end{flushright}


\begin{thebibliography}{100}
\bibitem{KSV} A.~Kitaev, A.~Shen, and M.~Vyalyi
{\it ``Classical and quantum computation"},
Graduate Studies   in Mathematics, Vol.~{\bf 47},
American Mathematical Society (2002).

\bibitem{KKR} J.~Kempe, A.~Kitaev, and O.~Regev
{\it ``The complexity of the local Hamiltonian problem"},
quant-ph/0406180.

\bibitem{APT} B.~Aspvall, M.~Plass, and R.~Tarjan  {\it ``A linear-time algorithm for
testing the truth of certain quantified boolean formulas"},
Info.~Proc.~Letters, Vol.~{\bf 8}, Iss.~3, p.~121-123 (1979).

\bibitem{ADKLLR}
 D.~Aharonov, W. van~Dam, J.~Kempe, Z.~Landau, S.~Lloyd, and O.~Regev
{\it ``Adiabatic Quantum Computation is Equivalent to Standard Quantum Computation"},
Proceedings of 45th FOCS (2004),
quant-ph/0405098.

\bibitem{AHU} A.~Aho, J.~Hopcroft, and J.~Ullman  {\it  ``The Design and
Analysis of Computer Algorithms''}, Addison-Wesley Reading, Massachusetts
(1974).


\bibitem{BP94} D.~Bini and V.Y.~Pan  {\it ``Polynomial and matrix
computation''}, Vol.~{\bf 1}, Birkh\"auser, Boston, Cambridge, MA (1994).

\bibitem{DV} David DiVincenzo, private communication.

\bibitem{Barenco}
A.~Barenco,
C.~H.~Bennett,
R.~Cleve,
D.~DiVincenzo,
N.~Margolus,
P.~Shor,
T.~Sleator,
J.~Smolin,
H.~Weinfurter
{\it ``Elementary gates for quantum computation"},
Phys.~Rev.~{\bf A52}, p.~3457 (1995).

\bibitem{Fenner} S.~Fenner, F.~Green, S.~Homer, R.~Pruim
{\it ``Determining Acceptance Possibility for a Quantum Computation is Hard for the Polynomial Hierarchy"},
quant-ph/9812056.




\end{thebibliography}
\end{document}